\documentclass[pdflatex,sn-nature]{sn-jnl}

\usepackage{graphicx}%
\usepackage{multirow}%
\usepackage{amsmath,amssymb,amsfonts}%
\usepackage{amsthm}%
\usepackage{mathrsfs}%
\usepackage[title]{appendix}%
\usepackage{xcolor}%
\usepackage{textcomp}%
\usepackage{manyfoot}%
\usepackage{booktabs}%
\usepackage{algorithm}%
\usepackage{algorithmicx}%
\usepackage{algpseudocode}%
\usepackage{listings}%

\usepackage{comment}

\theoremstyle{thmstyleone}%
\theoremstyle{thmstyletwo}%

\theoremstyle{thmstylethree}%

\begin{document}

\title[The revolution in strong lensing discoveries from Euclid]{The revolution in strong lensing discoveries from Euclid}


\author*[1]{\fnm{Natalie E. P.} \sur{Lines}}\email{natalie.lines@port.ac.uk}
\author[1]{\fnm{Tian} \sur{Li}}
\author[1]{\fnm{Thomas E.} \sur{Collett}}
\author[2]{\fnm{Philip} \sur{Holloway}}
\author[3]{\fnm{James W.} \sur{Nightingale}}
\author[4,1]{\fnm{Karina} \sur{Rojas}}
\author[2]{\fnm{Aprajita} \sur{Verma}}
\author[5,6]{\fnm{Mike} \sur{Walmsley}}

\affil[1]{\orgdiv{Institute of Cosmology and Gravitation}, \orgname{University of Portsmouth}, \orgaddress{\street{Burnaby Road}, \city{Portsmouth}, \postcode{PO1 3FX}, \country{UK}}
}

\affil[2]{\orgdiv{Department of Physics}, \orgname{Oxford University}, \orgaddress{\street{Keble Road}, \city{Oxford}, \postcode{OX1 3RH}, \country{UK}}}

\affil[3]{\orgdiv{School of Mathematics, Statistics and Physics}, \orgname{Newcastle University}, \orgaddress{\street{Herschel Building}, \city{Newcastle-upon-Tyne}, \postcode{NE1 7RU}, \country{UK}}}

\affil[4]{\orgdiv{School of Engineering}, \orgname{University of Applied Sciences and Arts of Northwestern Switzerland}, \orgaddress{\street{Bahnhofstrasse 6}, \city{Windisch}, \postcode{5210}, \country{Switzerland}}}

\affil[4]{\orgdiv{School of Engineering}, \orgname{University of Applied Sciences and Arts of Northwestern Switzerland}, \orgaddress{\street{Bahnhofstrasse 6}, \city{Windisch}, \postcode{5210}, \country{Switzerland}}}

\affil[5]{\orgdiv{David A. Dunlap Department of Astronomy \& Astrophysics}, \orgname{University of Toronto}, \orgaddress{\street{50 St George Street}, \city{Toronto}, \postcode{M5S 3H4}, \country{Canada}}}

\affil[6]{\orgdiv{Jodrell Bank Centre for Astrophysics}, \orgname{University of Manchester}, \orgaddress{\street{Oxford Road}, \city{Manchester}, \postcode{M13 9PL}, \country{UK}}}


\abstract{Strong gravitational lensing offers a powerful and direct probe of dark matter, galaxy evolution, and cosmology, yet strong lenses are rare: only 1 in roughly 10\,000 massive galaxies can lens a background source into multiple images. The European Space Agency’s \textit{Euclid} telescope, with its unique combination of high-resolution imaging and wide-area sky coverage, is set to transform this field. In its first quick data release, covering just 0.45\% of the full survey area, around 500 high-quality strong lens candidates have been identified using a synergy of machine learning, citizen science, and expert visual inspection. This dataset includes exotic systems such as compound lenses and edge-on disk lenses, demonstrating \textit{Euclid}’s capacity to probe the lens parameter space. The machine learning models developed to discover strong lenses in \textit{Euclid} data are able to find lenses with high purity rates, confirming that the mission’s forecast of discovering over 100\,000 strong lenses is achievable during its 6-year mission. This will increase the number of known strong lenses by two orders of magnitude, transforming the science that can be done with strong lensing.}

\maketitle

\section*{Introduction} \label{sec:Introduction}

\begin{figure}[H]
    \centering
    \includegraphics[width=0.9\linewidth]{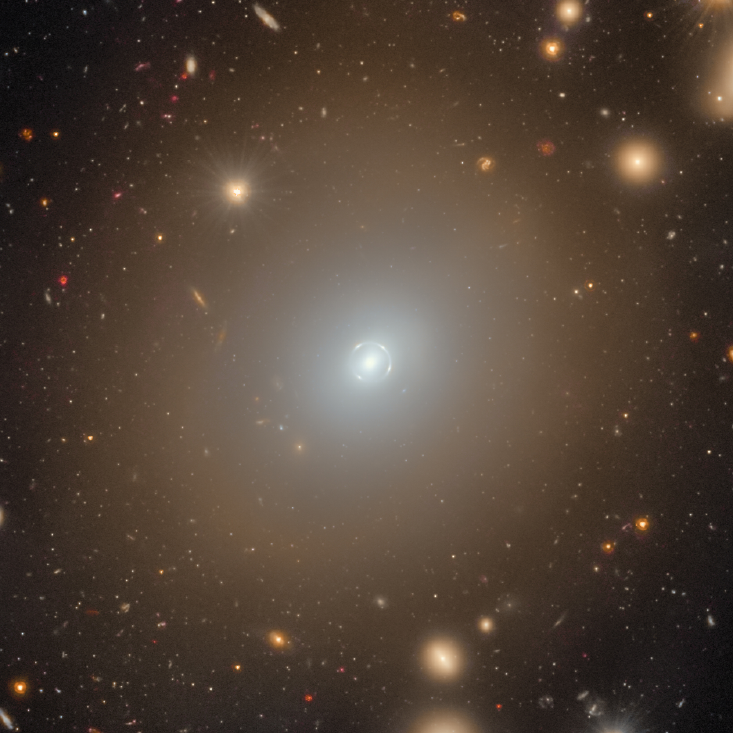}
    \caption{\textbf{An Einstein ring embedded in the nearby galaxy NGC 6505 imaged by \textit{Euclid}.} The image is created with the luminosity of the $I_{\mathrm{E}}$ band, while the red, green, and blue colours correspond to the $H_{\mathrm{E}}$, $J_{\mathrm{E}}$, and $Y_{\mathrm{E}}$ bands respectively.}
    \label{fig:ngc6505}
\end{figure}

Gravitational lensing is a striking demonstration of general relativity in action: massive galaxies deform space-time so severely that they distort light from background sources.
When a background object is nearly perfectly aligned with a massive foreground object, the distortion becomes so extreme that light takes multiple paths around the gravitational well, manifesting as multiple images in arcs or \textit{Einstein rings}.
The formation of multiple images marks the transition from weak gravitational lensing, which subtly distorts the average alignment of galaxies, to strong gravitational lensing, where the effects are prominent in individual systems \citep{zwicky1937, einstein1936}. An example of an Einstein ring, discovered with and imaged by \textit{Euclid} \citep{ngc6505}, is shown in Fig. \ref{fig:ngc6505}.
Strong lensing is a unique probe as it is sensitive to the total mass profile of a galaxy. It is a valuable tool for studying the poorly understood components of our Universe: dark matter and dark energy. Dark matter shapes the mass distribution of the foreground `lens' galaxy, while dark energy influences the expansion history of the Universe; together they affect the light from background `source' galaxies. Strong lensing provides insights into structure formation, probes galaxy properties at different redshifts, and enables independent measurements of key cosmological parameters.

However, a large sample of strong lenses is crucial for robust inference of different properties of galaxies. With thousands of lenses, it becomes possible to tightly constrain properties such as the mass density profile of galaxies \citep{shajib2021} and clusters \citep{newman2015}, the stellar initial mass function \citep{sonnenfeld2021}, the stellar-to-halo mass relation at lower mass scales \citep{wang2025}, the $M$–$\sigma$ relation of supermassive black holes \citep{Nightingale2023, cosmichorseshoebh}, cosmic shear \citep{hogg2023, birrer2018, etherington2024}, and the dark energy equation of state \citep{tian10kdspls}. Larger samples also enhance the likelihood of detecting rare phenomena: so far, only a few strong lenses have indicated the presence of dark matter subhaloes \citep{vegetti2010, vegetti2012,hezaveh2016}, and they have been found to be in tension with the cold dark matter model \citep{Minor:2020hic, Ballard:2023fgi, Despali:2024ihn, Enzi2024}. Large samples will also allow the discovery of more exotic strong lenses \citep{exoticlenses}, rarer lenses such as late-type disk galaxies \citep{edgeons1, edgeons2}, and enable detailed studies of the magnified source galaxies that are otherwise difficult to observe.
Furthermore, strong lensing can provide a powerful, independent method for measuring the Hubble constant, but requires a variable source such as supernovae or quasars \citep{holicow13}. Lensed supernovae can provide particularly tight constraints on the Hubble constant, but the combined rarity of strong lenses and supernovae makes such events very scarce.

\begin{figure} [H]
    \centering
    \includegraphics[width=1\linewidth]{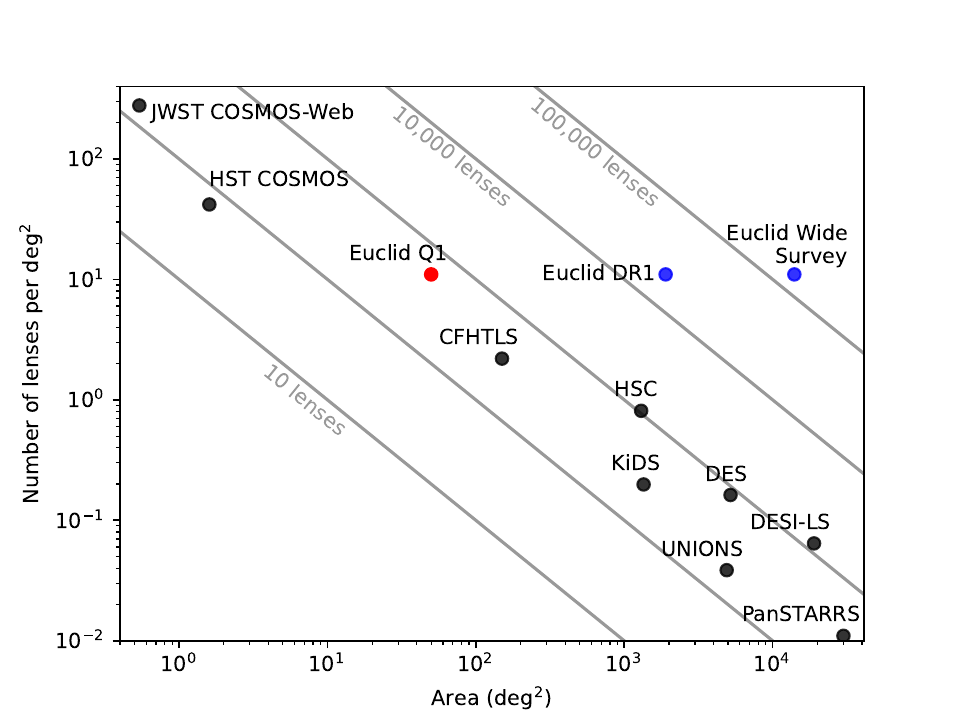}
    \caption{\textbf{The density of strong lenses against the survey area for different lens searches.} The number of lenses found in the Euclid Quick Data Release 1 (Q1, red), is extrapolated to Euclid Data Release 1 (DR1) and the full Euclid Wide Survey (blue), and compared with lens searches in other data (black). This includes the James Web Space Telescope (JWST) COSMOS-Web \citep{jwstcosmosweb, cowls2}, Hubble Space
    Telescope (HST) COSMOS \citep{faureFirstCatalogStrong2008}, Canada–France–Hawaii Telescope Legacy Survey (CFHTLS) \citep{Gavazzi2014},  Hyper Suprime-Cam (HSC) \citep{sonnenfeldSurveyGravitationallylensedObjects2018,sonnenfeldSurveyGravitationallylensedObjects2020a,Canameras2021,shu22_HOLISMOKES8, wongSurveyGravitationallyLensed2022,jaelani24,    schuldt_holismokes_2025, schuldtEtAl25b}, Dark Energy Survey (DES) \citep{jacobsExtendedCatalogGalaxyGalaxy2019,rojas_search_2022,gonzalez_discovering_2025}, Dark Energy Spectroscopic Instrument Legacy Surveys (DESI-LS) \citep{huang_finding_2020,huang_discovering_2021,storferNewStrongGravitational2024}, Ultraviolet Near-Infrared Optical Northern Survey (UNIONS) \citep{savaryStrongLensingUNIONS2022,edgeons2}, Kilo-Degree Survey (KiDS) \citep{liHighqualityStrongLens2021} and Panoramic Survey Telescope and Rapid Response System (PanSTARSS) \citep{Canameras2020}.
    Diagonal grey lines represent constant total numbers of lenses by order of magnitude. The Euclid Wide Survey is expected to contain a number of lenses two orders of magnitude larger than the number of currently known lenses.}
    \label{fig:lens-surveys}
\end{figure}

Unfortunately, strong lensing is extremely rare. Roughly 1 in 10\,000 massive galaxies are able to strongly lens another background galaxy, with the typical radius of an Einstein ring (Einstein radius) measuring just 0.4$^{\prime\prime}$ \citep{collett15}. Therefore, the ideal instrument for detecting strong lenses must have two key attributes: high angular resolution (to clearly differentiate lensed sources from the lens galaxy) and a large sky coverage (to compensate for the rarity of these systems). This makes the \textit{Euclid} space telescope a perfect telescope for strong lens discovery. Over \textit{Euclid}'s 6-year mission, it will survey a third of the whole sky with an angular resolution of 0.16$^{\prime\prime}$ \citep{EuclidSkyOverview}, revealing a wealth of strong lens systems that were previously undetectable. Although there have been many large-area surveys from Earth, atmospheric distortion typically restricts the resolution of ground-based telescopes to around 1$^{\prime\prime}$, limiting our ability to detect the bulk of the strong lens population.
As such, these ground-based surveys have only been able to discover a few thousand lens candidates in total, with the additional caveat that ground-based imaging makes it hard to confidently discern lenses from non-lenses \citep{q1mike}. In contrast, \textit{Euclid} is expected to discover over 100\,000 strong lenses in its lifetime -- two orders of magnitude more than currently known \citep{collett15}.
Figure \ref{fig:lens-surveys} presents an overview of many of the largest lens surveys to date with the corresponding areas and lens densities, demonstrating the extent to which \textit{Euclid} will revolutionise strong lensing.

\section*{The challenge} \label{sec:The-challenge}

\begin{figure} [H]
    \centering
    \includegraphics[width=1\linewidth]{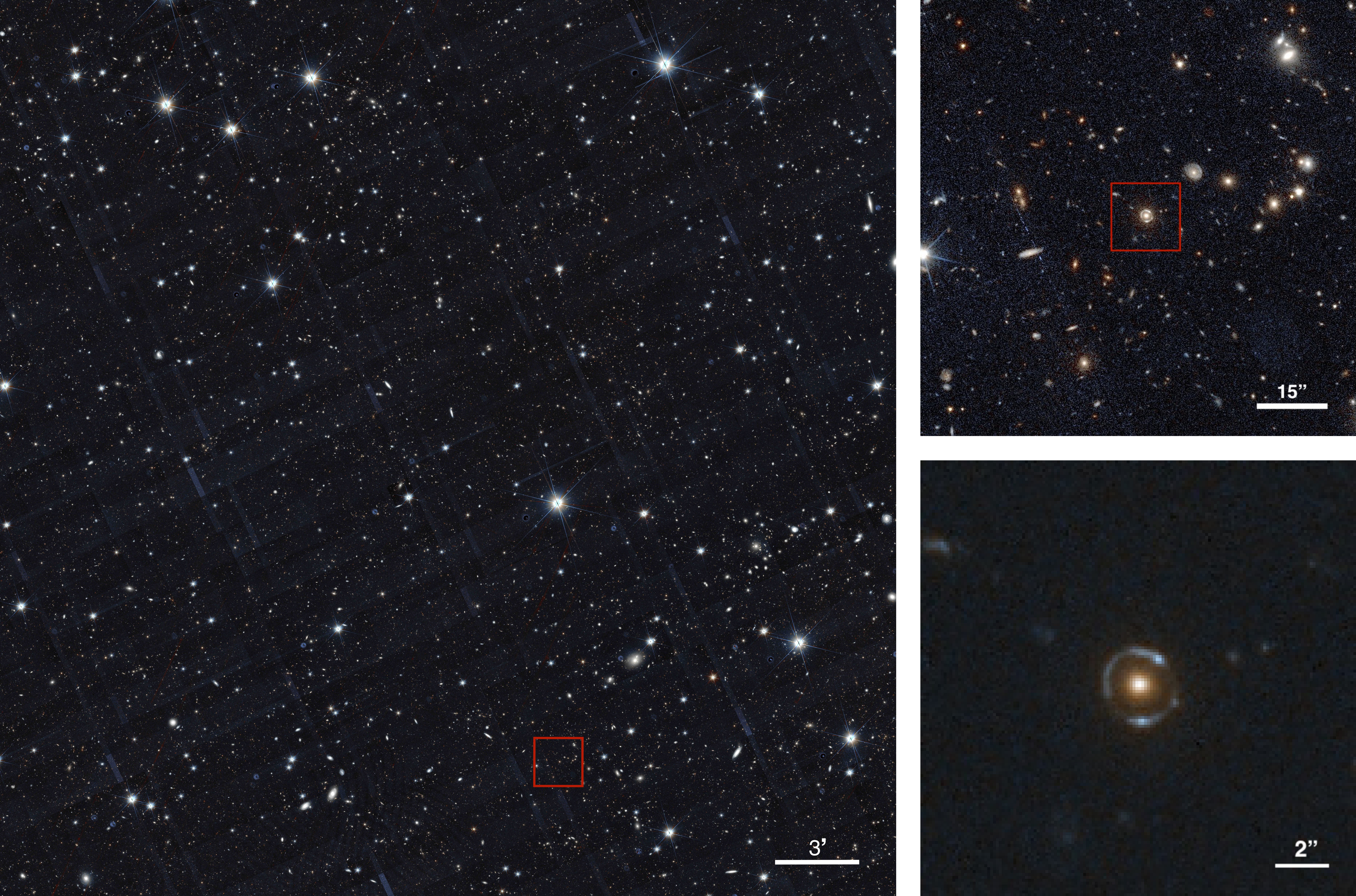}
  \caption{\textbf{An example strong lens found in a \textit{Euclid} tile.} Left: a full Euclid tile, corresponding to an area of 0.28 deg$^2$. Each tile typically contains one or two strong lenses. Top right: zoom-in of the region in the red box in the left image, where a strong lens can be spotted in the centre. Bottom right: zoom-in of the region in the red box in the top right image. The images are created with the luminosity of the $I_{\mathrm{E}}$ band, and are coloured using the $I_{\mathrm{E}}$ and $Y_{\mathrm{E}}$ bands.}
  \label{fig:full-tile}
\end{figure}

Figure \ref{fig:full-tile} demonstrates the scale of the needle-in-a-haystack problem that astronomers face: a single \textit{Euclid} tile contains tens of thousands of sources but only one or two strong lenses.
Searches for strong lenses have evolved from the first serendipitous discovery in 1979 \citep{walsh_0957_1979} to thousand-volunteer citizen science efforts (through projects such as `Space Warps' hosted on the Zooniverse platform \citep{Marshall2016,more_space_2016,garvin_hubble_2022}) and AI-driven scans of entire surveys \citep[e.g.][]{jacobsExtendedCatalogGalaxyGalaxy2019,rojas_search_2022, Canameras2020, Canameras2021, storferNewStrongGravitational2024, gonzalez_discovering_2025}. This evolution in search methods has been governed by necessity: bigger surveys resolve vastly more galaxies while the number of images professional astronomers can visually inspect remains roughly constant. Professional astronomers provide the final vetting of promising lens candidates, but it is impossible for an entire survey to be visually inspected by strong lensing experts, even with pre-screening by citizen scientists.
The largest visual inspections projects undertaken by both experts \citep{Jackson2008} and citizen scientists \citep{more_space_2016} have been limited to hundreds of thousands of images. In contrast, the full Euclid Wide Survey will image 1.5 billion galaxies, making it necessary to enlist the help of AI.

The main challenge for AI lens-finders is false positives: the rarity of strong lensing means that even a 99\% accurate AI classifier would return 100 false positives per true lens.
Spiral and ring galaxies can often confuse AI models, and even chance alignment of galaxies can resemble lensed arcs while being not much rarer \citep{q1natalie}.
The immense variety of objects in the Universe makes it difficult to train networks on a truly representative dataset, and when faced with something unfamiliar, a network could classify it either as `lens' as `non-lens'.
Strong lenses are also diverse, and the known examples are too few and heterogeneous to train robust machine learning models.
To compensate, researchers rely on simulations. However, machine learning models can fixate on minor details (such as noise patterns or point spread functions) and making simulations that perfectly replicate real telescope images is challenging: 
despite extensive efforts in developing lens simulations, discrepancies between lens-finding performance evaluated on simulations and real lenses persist.
It was therefore uncertain what proportion of \textit{Euclid}'s strong lenses could be extracted from the sea of stars and galaxies it will observe.

\section*{Euclid Q1} \label{sec:Euclid-Q1}

The \textit{Euclid} Q1 dataset allows the success of the lens searches in \textit{Euclid}'s early release observations to be built on and scaled up to the first 0.45\% of the full survey area. The Q1 data cover 63 deg$^2$ and include 30 million sources, posing a substantial challenge for lens finding.
To uncover strong lenses amongst this dataset, the most effective strategy combined all three lens-finding approaches: machine learning, citizen science, and expert visual inspection.

\begin{figure} [H]
    \centering    \includegraphics[width=1\linewidth]{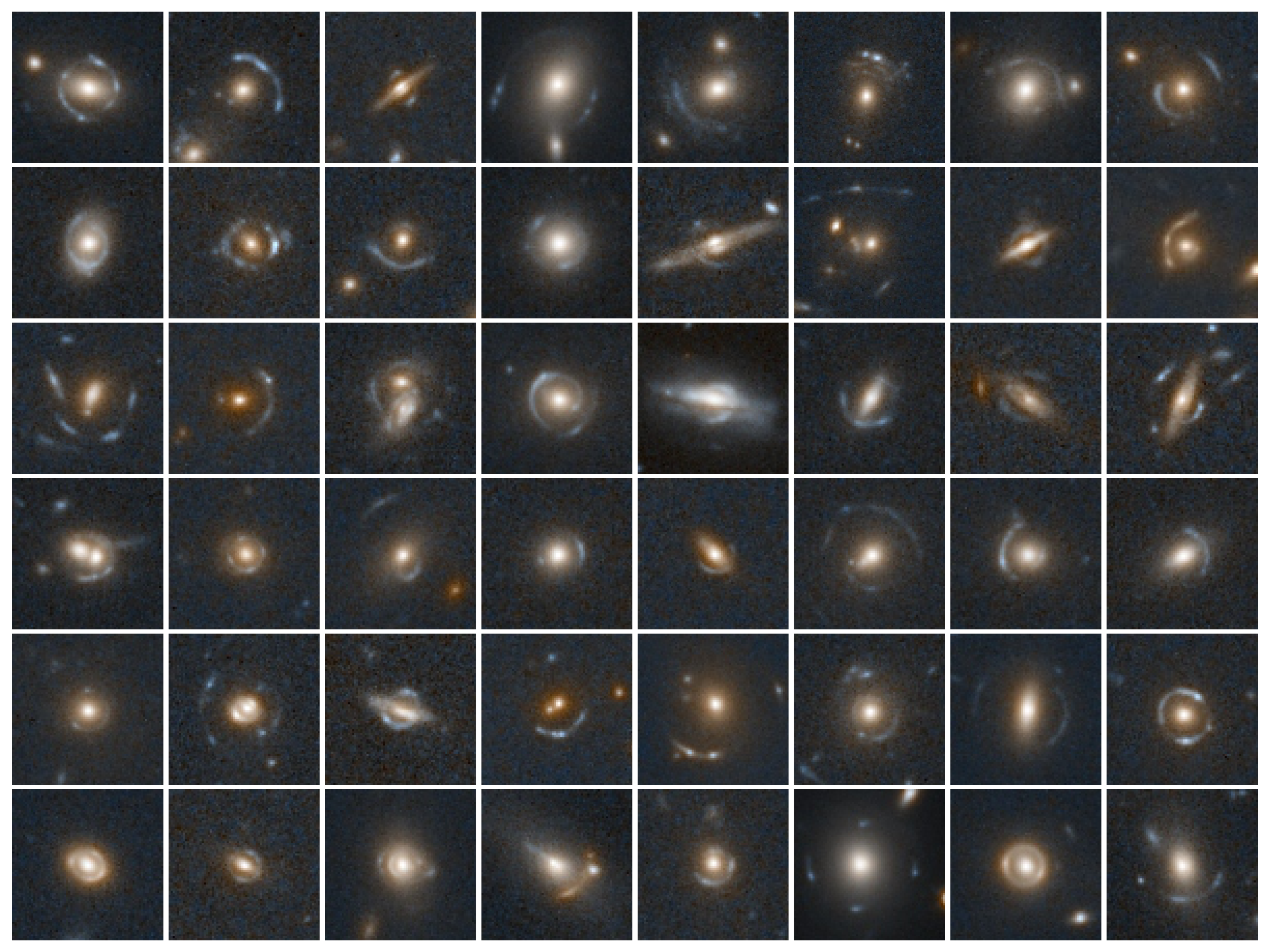}
    \caption{\textbf{Examples of strong lenses discovered in \textit{Euclid} Q1.} The images, which represent fewer than 10\% of the discovered lenses, are created with the luminosity of the $I_{\mathrm{E}}$ band and are coloured using the $I_{\mathrm{E}}$ and $Y_{\mathrm{E}}$ bands. Each cutout is $10^{\prime\prime}\times10^{\prime\prime}$.
    }
    \label{fig:cool_lenses}
\end{figure}

Five different machine learning models were trained to identify strong lenses in Q1, spanning a range of architectures including convolutional neural networks and vision transformers. Some were trained from scratch, while others built on pre-trained foundation models. The training data consisted primarily of simulated galaxy-galaxy lenses, created by adding lensed features to real \textit{Euclid} galaxy images to mimic real lenses as much as possible. To increase robustness, a diverse set of false positives that was collected from early visual inspections and morphologically classified was also included in training \citep{q1karina}.
The five models ranked around one million Q1 objects that passed an initial screening cut designed to exclude objects unlikely to be lens galaxies. 
As machine learning models are not perfect at recognising lenses, the most effective strategy was to visually inspect as many high-ranking candidates as possible. To do this, the visual inspection was crowd-sourced using a Space Warps project. This enabled 100\,000 cutouts to be reviewed. Promising lens candidates identified by citizen scientists were then graded by strong lensing experts, resulting in a catalogue of 500 high-quality strong lenses, almost all of which were new \citep{q1mike}. A selection of just a few of these new strong lenses is shown in Fig. \ref{fig:cool_lenses}.

The best performing machine learning model in Q1 was a fine-tuned version of Zoobot, a model that had been pre-trained on general galaxy morphologies from other astronomical surveys \citep{Walmsley2023zoobot}. This pre-training exposed the network to the diversity of galaxies in the Universe, enabling it to more rapidly adapt to identifying features of strongly lensed galaxies. Zoobot was first applied to lens-finding in \textit{Euclid}'s early release observations where it was found to perform very well compared to other machine learning approaches \citep{rubyero}. Building on this for Q1, it was found that varying the depth to which the network was fine-tuned made a significant impact on performance. Additionally, the importance of having diverse training data was highlighted: from inspecting the non-lenses with high predicted lens probabilities, it was clear that one key class of false positives was chance alignment of elongated galaxies around elliptical galaxies, mimicking lensed arcs. False positives that have provided problems for previous machine learning networks such as spiral and ring galaxies were much less prevalent amongst the false positives. Since serendipitous alignment of galaxies resembling lenses is much less frequent than ring galaxies, this resulted in a higher-purity sample of strong lenses, enabling more lenses to be discovered quickly during visual inspection.

\begin{figure*}
\centering
\includegraphics[width=1\linewidth]{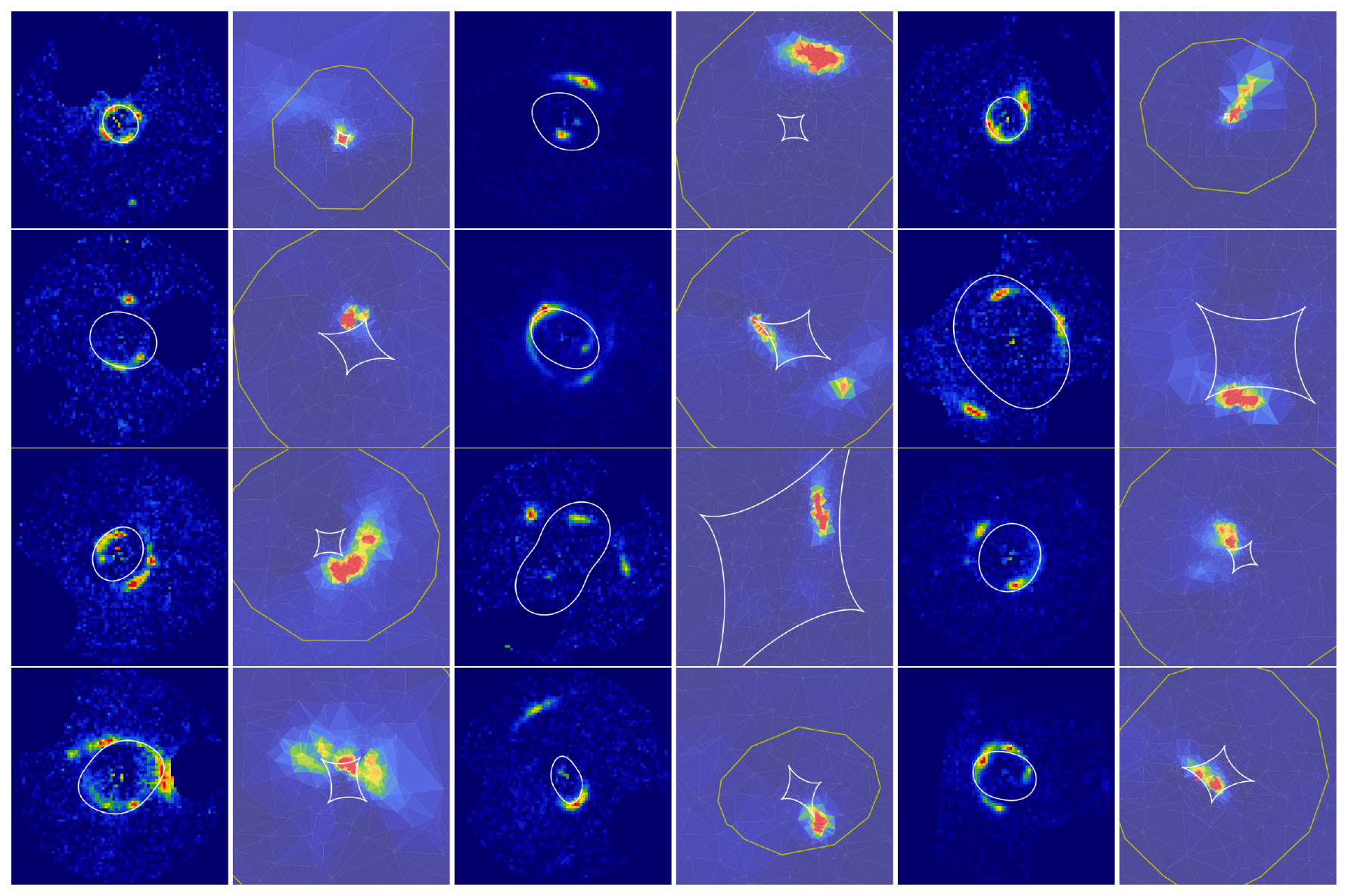}
\caption{\textbf{A mosaic of 12 Euclid lensed galaxies and their corresponding reconstructed sources.} For each lens, we show the foreground-subtracted $I_{\mathrm{E}}$ image (left) and the corresponding source-plane reconstruction (right) produced using \texttt{PyAutoLens}. The lenses were selected to illustrate a diverse range of Einstein radii, lensing configurations and source morphologies. Foreground light was subtracted using a multi-Gaussian expansion model. The mass model assumes a singular isothermal ellipsoid profile with external shear, and the background source is reconstructed on an adaptive Voronoi mesh. White and yellow curves denote tangential and radial critical curves in the image-plane and caustics in the source-plane, respectively. All image-plane cutouts are 3.5$^{\prime\prime}$
across and the patches that contain no noise are where additional line-of-sight galaxies have been removed before lens modelling. Blue regions correspond to areas of low intensity and red regions to areas of high intensity. All source-plane reconstructions span 0.9$^{\prime\prime}$ around the same centres as the images.
}
\label{figure:Lensmodel}
\end{figure*}

Discovering lenses at this rate means that the automation of lens modelling is necessary. Using the lens modelling software \texttt{PyAutoLens} \citep{nightingalePyAutoLensOpenSourceStrong2021}, automated modelling was successfully performed on over 300 of the lens systems, indicating they are almost certainly lenses \citep{q1mike}. Figure \ref{figure:Lensmodel} presents a mosaic of 12 representative lens model fits and source reconstructions, selected to span a range of Einstein radii, lens configurations, and source morphologies. The models adopt a singular isothermal ellipsoid mass profile with external shear, and successfully map the multiple images in the image plane to coherent regions in the source plane. Several sources exhibit complex, multicomponent morphologies, suggestive of ongoing mergers.

Most of these high-quality lens candidates have configurations with no other reasonable astrophysical explanation, although some of the lower-quality candidates require spectroscopic measurements to verify that the multiple images originate from the same source. Most lensing science also depends on robust redshift measurements of both the lens and source galaxies.
While \textit{Euclid} is equipped with spectroscopic capabilities, it is unable to provide accurate redshift measurements for most of the elliptical lens galaxies as the exposures are too shallow to reveal absorption lines in the continuum. This necessitates follow-up with other instruments, though this can also provide a challenge.

\subsection*{500 strong lenses found} \label{ssec:500-strong-lenses-found}

\begin{figure} [H]
  \centering
    \includegraphics[width=1\linewidth]{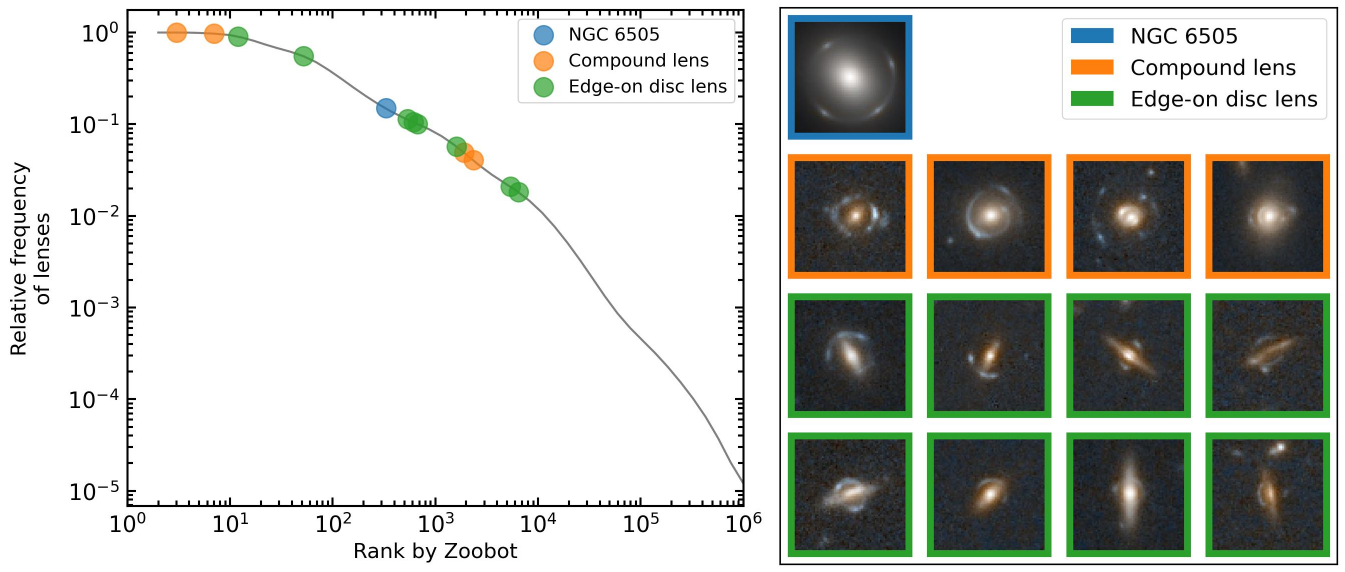}
  \caption{\textbf{The different types of strong lenses found by the fine-tuned Zoobot model.} {Left}: The density of lenses found in Q1 as a function of Zoobot rank is shown. Scattered points mark the ranks of lenses of particular interest, including the only discovered lens around an NGC galaxy, the four compound lenses, and eight of the edge-on disk lenses. {Right}: Images of the interesting lenses corresponding to objects that are plotted on the left, ordered by their Zoobot rank. The cutouts are $10^{\prime\prime}\times10^{\prime\prime}$ and are created with the luminosity of the $I_{\mathrm{E}}$ band and colours from the $I_{\mathrm{E}}$ and $Y_{\mathrm{E}}$ bands.}
  \label{fig:lens-distribution}
\end{figure}

This catalogue of lenses in Q1 represents the largest sample of strong lenses with high-resolution imaging to date.
By combining wide-area coverage with space-based resolution, \textit{Euclid} is enabling the detection of strong lenses that were previously missed, opening new windows into lens populations that have historically been under-represented.
The Q1 sample alone notably contains four new compound lenses -- lens systems with a single galaxy lensing multiple source galaxies -- nearly doubling the total number known \citep{Q1tian}. Additionally, around 40 new edge-on disk galaxy lenses were discovered, a rare class of lens due to their typically lower masses, which are particularly interesting probes of dark matter. 
These discoveries are also noteworthy as the machine learning models had not been trained to find such lens systems, demonstrating the power and adaptability of the networks. Figure \ref{fig:lens-distribution} shows the distribution of how these lenses were ranked by the best performing model, along with images of some of the rarer lenses.

Although visually inspecting only 10\% of the full dataset meant there will be unknown unknowns, a statistical analysis based on a representative sample of the Q1 data suggests that almost all of the highest-quality strong lenses were found \citep{q1natalie}. Without AI, the only way of confidently finding close to 100\% of the lenses in a sample would be to visually inspect every image, which is not feasible for the quantities of data that \textit{Euclid} will collect. Although there have been significant developments in AI in recent years, finding strong lenses had remained an open challenge, and discovering even a large fraction of these strong lenses was not guaranteed with the previously available methods. 
The success of the Q1 search marks a major step forwards and provides strong evidence that these machine learning models are ready to meet the demands of analysing upcoming \textit{Euclid} data.

\section*{Future} \label{sec:Future}
This is just the beginning for \textit{Euclid}: Q1 represents only one week of \textit{Euclid} data, yet it already rivals the largest strong lens catalogues previously available.
Scaling up the results from Q1 by area, \textit{Euclid} is on track to discover 15\,000 strong lenses in the first major data release (DR1, available publicly in late 2026), and around 110\,000 by the end of the full mission. This will include around 1000 compound lenses \citep{Q1tian} and 2500 subhalo detections \citep{oriordanSensitivityStrongLensing2023}.
Unfortunately, scaling up the lens finding from Q1 is not without challenges. Running a machine learning algorithm on such large quantities of data is feasible, given that it took less than an hour on a sufficiently powerful computing cluster for Q1. However, it took citizen scientists a few weeks for the Q1 visual inspection, and directly scaling this up means that inspecting the same proportion of the full mission would take decades. Fortunately, inspecting this many images is not necessary to discover most of the lenses.
By ranking candidates with the best-performing Q1 machine learning model, inspecting the top million images in the full \textit{Euclid} survey -- an achievable number over the 6-year period -- will result in 75\,000 new strong lenses \citep{q1natalie}.
This represents a transformative sample size, large enough to enable a wide range of scientific studies with strong lensing that were previously unachievable.
This number provides a conservative lower limit, as combining multiple machine learning models boosts lens-finding performance \citep{Q1phil}. 
In addition, preliminary retraining efforts using the Q1 lenses and false positives have already produced a model that can discover the first quarter of the lenses in a Q1-size sample with a purity of 94\%, doubling the purity compared to the previous version of the network. This level of purity demonstrates that the rate of improvement in machine learning performance can make up for limitations in the data volume that can be visually inspected. Given the improvement from retraining on the first 0.45\% of \textit{Euclid} data, there are likely to be further significant increases in AI lens-finding abilities.

With these numbers of strong lenses, we are entering an era where the challenge is no longer just finding lenses, but finding the most valuable ones. For example, while the number of compound lenses discovered in Q1 aligns with forecasts, relying on humans to flag interesting lens systems may be insufficient to uncover all compound lenses in the next data releases. Therefore, a tailored search for these specific lens systems is likely necessary. The galaxy-scale lens search in Q1 did not specifically target lensed quasars as their appearance is visually distinct from that of lensed galaxies, and therefore a dedicated search for these will also be needed. Forecasts for using strong lenses for time-delay cosmography show that the optimal approach is a synergy between \textit{Euclid} and the Vera Rubin Observatory's upcoming Legacy Survey of Space and Time, which will survey half the sky every three days \citep{lsst}. This joint programme is expected to deliver roughly seven cosmology-grade lensed type Ia supernovae per year, while simultaneously recording precise time-delay light curves for the lensed quasars. Provided that systematics can be accounted for, these datasets will enable tight measurements of the Hubble constant: 1.3 $\%$ precision from the lensed supernova sample and sub‑percent precision from the lensed quasar sample, both achievable within the coming decade \citep{ana2024, shajib25}. Follow-up observations of all the different types of \textit{Euclid} lenses will provide strong constraints on the equation of state of dark energy, with a predicted figure of merit of approximately 60 \cite{shajib25}. After decades of suffering from sparse data, \textit{Euclid} is driving a new era for strong lensing, greatly expanding its power as a precision tool for exploring the fundamental physics of our Universe.

\backmatter

\bmhead{Acknowledgements}
\textit{Euclid} is an ESA mission with major contributions by the
Euclid Consortium, consisting of more than 2000 scientists, engineers,
and technicians from 15 European countries, the USA, Canada, and
Japan.

\section*{Declarations}

\begin{itemize}
\item Competing interests:
The authors declare no competing interests.

\item Author contributions: N. L. led the structuring of the text, wrote the initial version of the manuscript, and produced Figures \ref{fig:full-tile}, \ref{fig:cool_lenses} and \ref{fig:lens-distribution}. T. L. produced Figure \ref{fig:ngc6505}. M. W. produced Figure \ref{fig:lens-surveys}. J. N. produced Figure \ref{figure:Lensmodel}. All authors contributed to the research discussed as well as reviewing the full text and contributing to the editing of the manuscript.

\end{itemize}

\bigskip





\bibliography{clean_bibliography}

\end{document}